\documentclass[prb,aps,twocolumn,amsmath,amssymb,superscriptaddress]{revtex4-1}

\usepackage{times}
\usepackage{textcomp}
\usepackage{graphicx}
\usepackage[caption=false]{subfig}
\usepackage{bm}
\allowdisplaybreaks 
\usepackage{braket}
\usepackage{lipsum}
\usepackage{amsmath}

\makeatletter

\newcommand{\Rmnum}[1]{\expandafter\@slowromancap\romannumeral #1@}
\makeatother

\usepackage[breaklinks]{hyperref}
\hypersetup{colorlinks=true, linkcolor=blue, citecolor=blue, filecolor=blue, urlcolor=blue}

\begin{document}

\title{Machine Learning Regression for Operator Dynamics}

\author{Justin A. Reyes}
    \affiliation{Department of Physics,University of Central Florida, Orlando, FL, 32816, USA}

\author{Sayandip Dhara}
    \affiliation{Department of Physics,University of Central Florida, Orlando, FL, 32816, USA}

\author{Eduardo R. Mucciolo}
    \affiliation{Department of Physics,University of Central Florida, Orlando, FL, 32816, USA}
    
\date{\today} 

\begin{abstract}
Determining the dynamics of the expectation values of operators acting
on quantum many-body systems is a challenging task. Matrix product
states (MPS) have traditionally been the "go-to" models for these
systems because calculating expectation values in this representation
can be done with relative simplicity and high accuracy. However, such
calculations can become computationally costly when extended to long
times. Here, we present a solution for efficiently extending the
computation of expectation values to long time intervals. We utilize a
multi-layer perceptron (MLP) model as a tool for regression on MPS
expectation values calculated within the regime of short time
intervals. With this model, the computational cost of generating
long-time dynamics is significantly reduced, while maintaining a high
accuracy. These results are demonstrated with operators relevant to
quantum spin models in one spatial dimension.
\end{abstract}


\maketitle

\section{Introduction}
\label{sec:introduction}

The accurate determination of expectation values for operators acting
on quantum many-body (QMB) systems at long times remains an open
problem. Much progress has been made for various specific systems of
interest, such as the Ising chain with a quenched transverse
field,\cite{Calabrese2011} or the Ohmic spin-boson model coupled to a
harmonic non-Markovian environment.\cite{Strathearn2018} However,
these developments have focused on systems where symmetries and
approximations can be exploited, analytic or exact diagonalization
methods can be used, or matrix product state algorithms can be
employed. Such approaches are either limited in their scope or quickly
become computationally demanding, particularly for systems in more
than one spatial dimension. This is particularly true for the standard
time-evolving block decimation (TEBD) \cite{Paeckel2019},
time-dependent density matrix renormalization group
($t$-DMRG),\cite{Vidal2004} and dynamic density matrix renormalization
group (DDMRG) algorithms.\cite{Jeckelmann2002}

Recent advances in machine learning models have offered new insights
and paved new pathways for modeling QMB systems, often providing
significant computational advantages over traditional
methods.\cite{Marcello2019,Melkinov2019,Soto2019} Motivated by these
successes, we investigate the advantages machine learning can provide
when computing the expectation values for operators acting on QMB
systems within the long-time regime.

Previous work utilizing machine learning techniques in QMB systems was
heavily focused on the use of restricted Boltzmann machines (RBMs) as
generative models of quantum
states.\cite{Nomura2017,Gao2017,Glasser2018} These are energy-based
models with an energy cost function given by
\begin{equation}
E(v,h) = -\sum_i{a_iv_i} - \sum_j{b_jh_j} -\sum_{i,j}{v_iW_{ij}h_j}.
\end{equation}
Here, $v = \lbrace v_i \rbrace$ and $h=\lbrace h_j \rbrace$ are the
visible and hidden layers of neurons in the RBM network,
respectively. Each visible (hidden) neuron can only take on the values
$\pm 1$ with an associated bias $a_i$ ($b_j$) and is fully connected
to the hidden (visible) layer by the weight matrix
$W$.\cite{Montufar2018} Given a spin-1/2 system, the probability
amplitude of a specific spin state, $\Psi_{\mathrm{RBM}} (v)$, can be
represented by the RBM by setting a spin configuration $v$ for the
visible layer and performing a summation over all hidden variables as
\begin{eqnarray}
\Psi_{\mathrm{RBM}} (v) & = & \sum_h{e^{-E(v,h)}} \\ & = &
\prod_i{e^{a_iv_i}}\prod_j{1 + e^{b_j + \sum_iv_iW_{ij}}}.
\end{eqnarray}
To obtain information about the full state of the system, it is
necessary to perform sampling over numerous spin configurations. While
this model is preferably suited to the determination of ground state
properties, after some modifications it has also been used to
determine dynamical properties of QMB
systems.\cite{Hartmann2019,Carleo2017,Hendry2019} Recently,
convolutional neural network have also been used to map input QMB spin
configurations to probability amplitudes.\cite{Schmitt2020} In spite
of the success of these approaches, these models still face similar
challenges as other computational methods, namely that accurately
representing the system state becomes computationally demanding as the
system size grows.

To circumvent the computational demands of representing (or sampling)
the full state of the system at any given time, we focus our attention
on the direct evolution of expectation values by breaking time into two
domains. For any given operator $\mathcal{O}$ acting on a quantum
system $\ket{\mathbf{\Psi} (t)}$, the expectation value of the
operator at any given time is given by
\begin{equation}
\label{eq:expectation_value}
\langle \mathcal{O} \rangle =
\bra{\mathbf{\Psi}(t)}\mathcal{O}\ket{\mathbf{\Psi}(t)}.
\end{equation}
By computing $\langle \mathcal{O}\rangle$ using matrix product state
(MPS) algorithms within the short-time domain, we shown that the
long-time expectation values can be determined with low computational
effort and good accuracy by utilizing a multi-layer perceptron (MLP)
as a tool for linear regression. It is noted that previous
extrapolation methods have been studied, but these have either focused
on constructing the wave function at each increment of
time~\cite{Tian2020} or implementing linear prediction methods to
$t$-DMRG spectral calculations.\cite{Barthel2009}

This paper is organized as follows: In Sec.~\ref{sec:mlp}, we review
the fundamentals of the multi-layer perceptron (MLP) model. In
Sec.~\ref{sec:tebd}, as a benchmark for algorithmic comparison, we
review the time-evolving block decimation algorithm in the context of
calculating operator dynamics. In Sec.~\ref{sec:regression}, we
describe the methodology involved in using the MLP for regression. In
Sec.~\ref{sec:results}, we demonstrate the computational advantage
gained by using MLP regression to determine operator dynamics for both
the Ising and the XXZ model. Finally, in Sec.~\ref{sec:discussion}, we
interpret these results and provide a framework for further
improvement and investigation.

\section{Multi-layer Perceptrons}
\label{sec:mlp}

\begin{figure}[tb]
\centering
\includegraphics[width=\columnwidth]{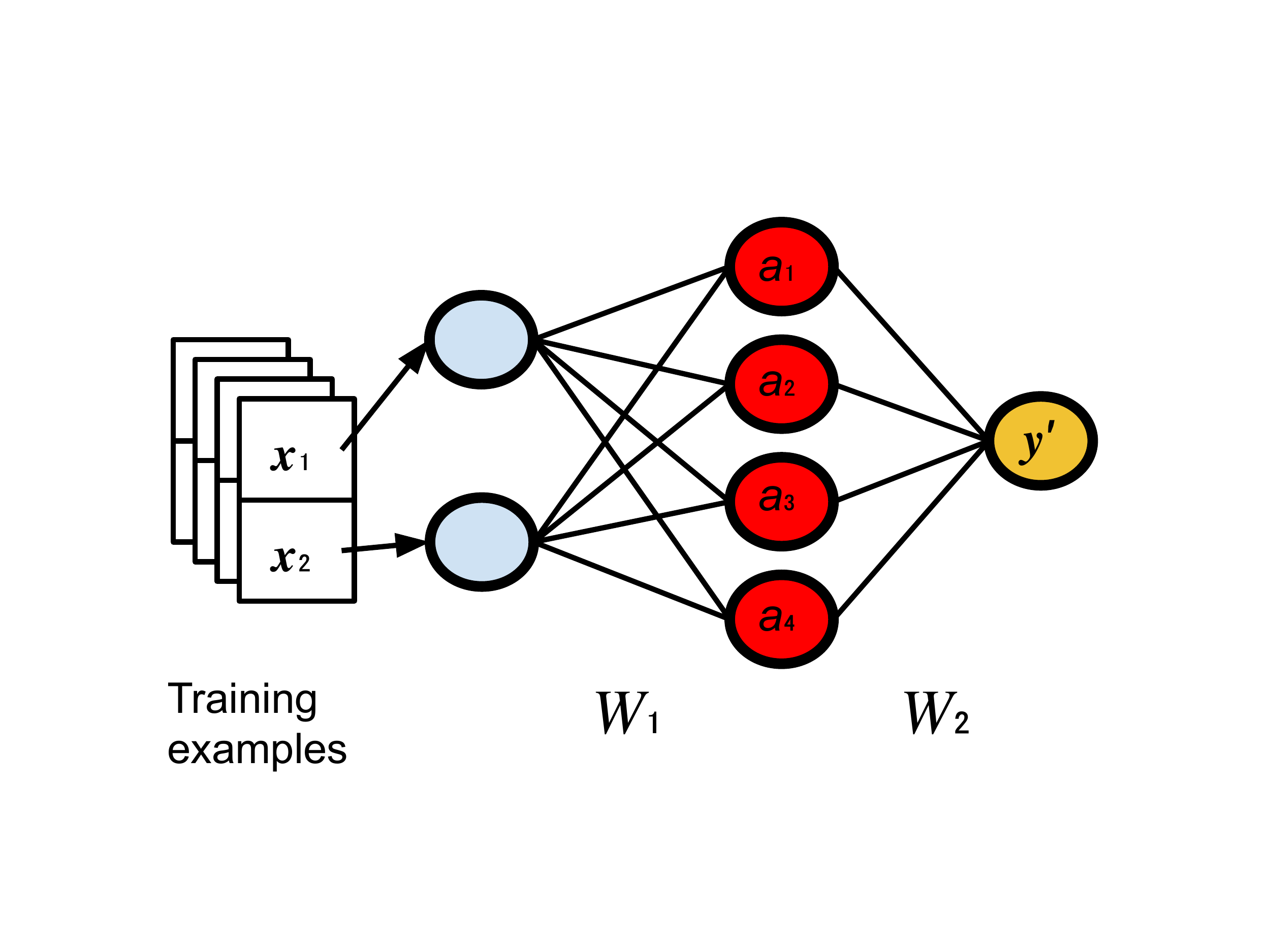}
\caption{A graphical representation of an example MLP, with input
  vectors $\vec{x}_n$ each having two elements for use as input to the
  first layer of neurons (blue). This input is propagated to the next
  layer (red) by interconnecting weights $W_1$ and finally sent to the
  output (yellow) with weights $W_2$.}
\label{fig:rho_mlp}
\end{figure}{}

\subsection{Architecture}

In machine learning, the MLP model is a ubiquitous tool for performing
classification tasks.\cite{Gardner1998} It is an input-output model
approximating the function
\begin{equation}
y' = W\cdot f(\mathbf{x}),
\end{equation}
where $f(\mathbf{x})$ is an activation function over a set of inputs
$\mathbf{x}$, $W$ is a weight matrix, and $y'$ is a guessed
classification label. This model is composed of $l$ sequential layers
of neurons $\lbrace a^{(i)}_{n_i}\rbrace$, where $0<i<l$, and $n_i$
specifies the number of neurons in a given layer $i$. Each neuron is
subject to the activation function $f$. Additionally, each layer of
neurons has a specified weight matrix $W^{(i)}$, which connects the
output from one layer to the input of the next. The components of each
$W^{(i)}$ are used as parameters for optimization of the
network.\cite{Gardner1998} Figure~\ref{fig:rho_mlp} provides a
schematic for a MLP with a single layer of neurons between the input
and final output.

\subsection{Supervised Learning}

To minimize the cost of the guessed label $y'$ generated by the MLP,
we provide an initial data set having $N$ elements,
$\lbrace\mathbf{x}_n,y_n\rbrace$, where $0<n<N$, for use in a training
protocol. This provision for training is characteristic of
\textit{supervised learning}.\cite{Kotsiantis2007} In this learning
procedure, each input vector $\mathbf{x}_n$ is accompanied by a
corresponding true classification label $y_n$. This label provides the
reference for a cost function $C(y_n',y_n)$ which measures the
distance between the current MLP output classification label $y_n'$
and the true classification label $y_n$.\cite{Kotsiantis2007} For our
specific purposes,xs we define $C(y_n',y_n)$ as
\begin{equation}\label{eq:cost}
C(y_n',y_n) = \frac{1}{N}\sum_n^N |y_n' - y_n |.
\end{equation}
Minimizing this cost function can be accomplished by any selection of
known optimizations algorithms. For our purposes, we focus on using a
stochastic gradient descent method.\cite{Zhang2004}

\section{Time Evolving Block Decimation}
\label{sec:tebd}

Before introducing the MLP regression algorithm, we provide a short
review of the TEBD algorithm so that computational comparisons to our
algorithm might be well understood. The TEBD algorithm facilitates the
time evolution (real or imaginary) of one-dimensional quantum systems
under local Hamiltonians.\cite{Paeckel2019} As such, it is naturally
expressed within the framework of matrix product states (MPS). The
time evolution is accomplished by generating and repeatedly applying
Suzuki-Trotter expansions of the time evolution operator
$\exp{(-i\mathcal{H} T)}$, up to any specified order. Given a
Hamiltonian with nearest-neighbor interactions and open boundary
conditions (OBC) over $N$ sites,
\begin{equation}\label{eq:loc_Ham}
\mathcal{H} = \sum_{i=1}^{N-1} H_{i,i+1},
\end{equation}
the second-order Suzuki-Trotter expansion of $\exp{(-i \mathcal{H}
  T)}$ for a small time-step $\delta > 0 $ is given as
\begin{eqnarray}
e^{-i T \mathcal{H}} & \approx & \left[ (e^{-\frac{i\delta }{2}
    H_{1,2}} e^{-\frac{i\delta }{2} H_{3,4}} \cdots e^{-\frac{i\delta
    }{2} H_{n-1,n}} ) \right. \nonumber \\ & &
  \left. \times\ (e^{-i\delta H_{2,3}} e^{-i\delta H_{4,5}} \cdots
  e^{-i\delta H_{n-2,n-1}}) \right. \nonumber \\ & &
  \left. \times\ (e^{-\frac{i\delta}{2}H_{1,2}}
  e^{-\frac{i\delta}{2}H_{3,4}} \cdots
  e^{-\frac{i\delta}{2}H_{n-1,n}}) \right]^{T/\delta}.
\end{eqnarray}
The sequential application of these operators demands that the MPS be
brought to canonical form (i.e., orthonormalizing the indices) after
every time step $\delta$.\cite{Vidal2004} Such a procedure involves
$O(poly(N)poly(D))$ steps, where $D$ is the maximum internal bond
dimension of the MPS. In the absence of truncation, $D$ grows
exponentially with both the system size and the evolution time and the
computational cost of this procedure quickly becomes intractable for
large systems and long times.
     
After the application of these operators, the final state of the
system is obtained as
\begin{equation}
\ket{\Psi (t=T)}  =e^{-i T \mathcal{H}} \ket{\Psi(t=0)}.    
\end{equation}
Typically, there are two sources of error in the TEBD framework. The
first comes from the truncation of the MPS bond dimensions during the
orthonormalization process. The other source of error arises during
the Suzuki-Trotter expansion, which for our purposes is taken to
second order. In this case, the error per time step is on the order
$O(\delta^3)$ resulting in an error over the total time interval on
the order of $O(\delta^2)$. In this paper, we choose to mitigate the
first source of error by performing minimal amounts of truncation on
the MPS (i.e., maintaining large bond dimensions). This is done to
ensure that the primary source of error arises from the Trotter-Suzuki
approximation itself.

\section{Machine Learning Regression}
\label{sec:regression}

\begin{figure}[tb]
\centering
\includegraphics[width=\columnwidth]{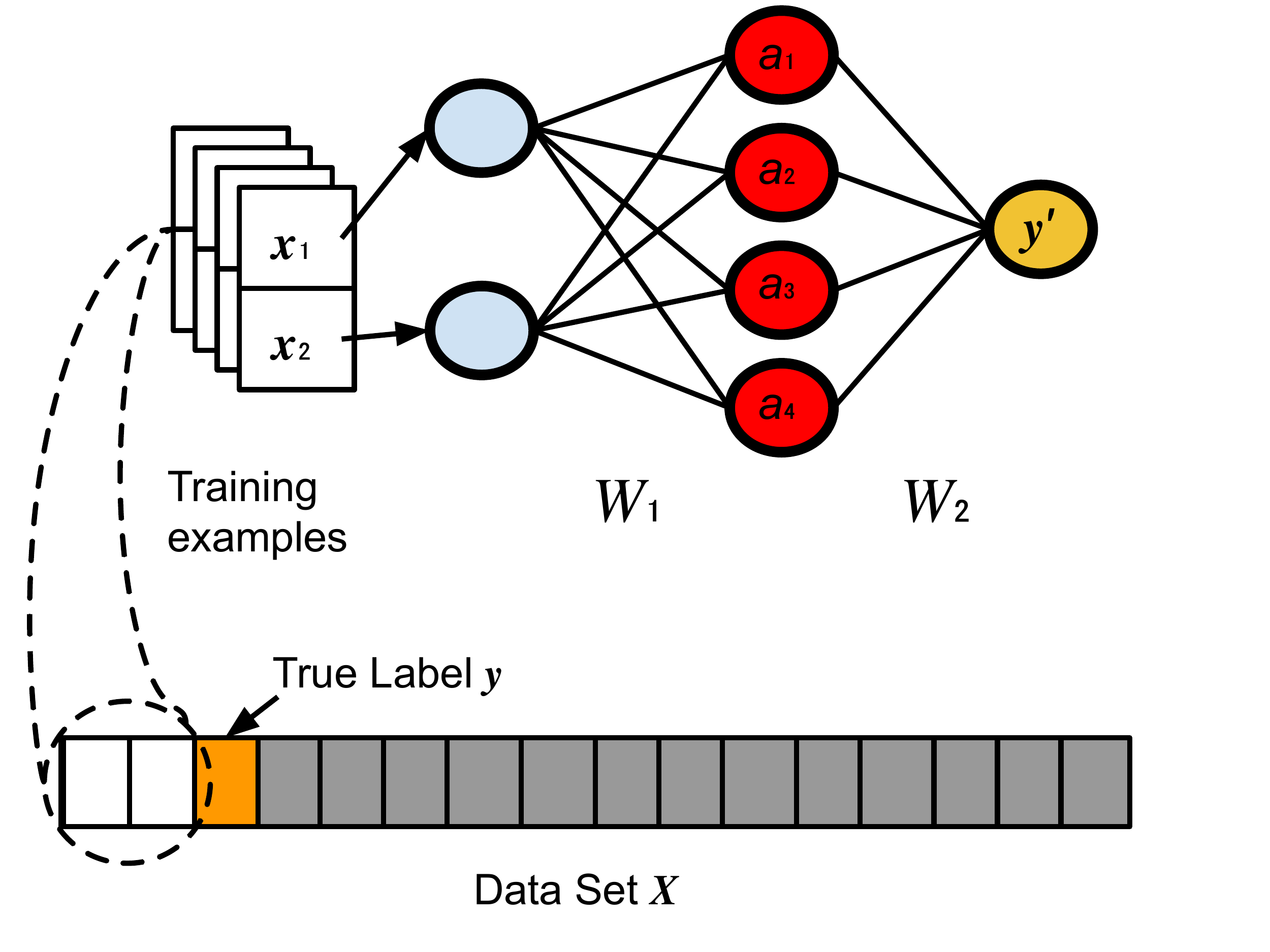}
\caption{An example diagram for MLP regression. The input data set $X$
  is decomposed into sets of training examples consisting of input
  vectors (white) and output values (yellow) selected from contiguous
  blocks in $X$. These training examples are fed into the MLP as
  shown. }
\label{fig:mlp_regression}
\end{figure}

In order to effectively model the evolution of operator expectation
values, we construct the MLP in a manner conducive to regression
rather than classification. Accomplishing this involves a few
specifications about the input-output pairs
$\lbrace\mathbf{x},y\rbrace$. We treat an input vector $\mathbf{X}$ as
being parameterized by time $t$ over a time interval $[0,\tau]$ so
that each element $X_i\in \mathbf{X}$ is labeled by a coordinate
$t_i$. The total time $\tau$ is partitioned into $m$ discrete time
intervals $\lbrace t_i | 0<i<m \rbrace$. From $\mathbf{X}$, each
input-output pair is constructed as follows. Starting from the first
element in $\mathbf{X}$ corresponding to time $t_0=0$, we select a
contiguous block of $p$ elements from $\mathbf{X}$ to form an input
vector $\mathbf{x} = \lbrace X_0, X_1, ... , X_p\rbrace$. We call this
block our \textit{training window}. The corresponding label for this
window is selected as the element $X_{p+1}$. To construct multiple
input examples for training, we shift the starting position of the
training window throughout $\mathbf{X}$ until the desired number of
examples is achieved. A diagram of this initialization procedure is
given in Fig.~\ref{fig:mlp_regression}.

In addition to constructing the input-output pairs in the
aforementioned manner, we choose to define our activation functions by
the linear unit, $f=x$. This activation allows us to effectively
propagate all of the input information through the network. This is in
contrast to the more commonly used rectified linear unit (ReLU),
$f=\max (0,x)$, which, depending on the values selected for the
weights, can suppress some information propagation through the network
by eliminating all negative values.\cite{Nwankpa2018} The ReLU
activation is useful when the MLP is used for classification over a
discrete set of positive valued labels. However, we select the linear
activation because our output values are continuous and include values
less than zero.

\vspace{-.5cm}

\section{Results}
\label{sec:results}

\begin{figure}[tb]
\captionsetup[subfigure]{labelformat=empty}
  \centering
    \subfloat[][]{
      \includegraphics[width=0.42\textwidth]{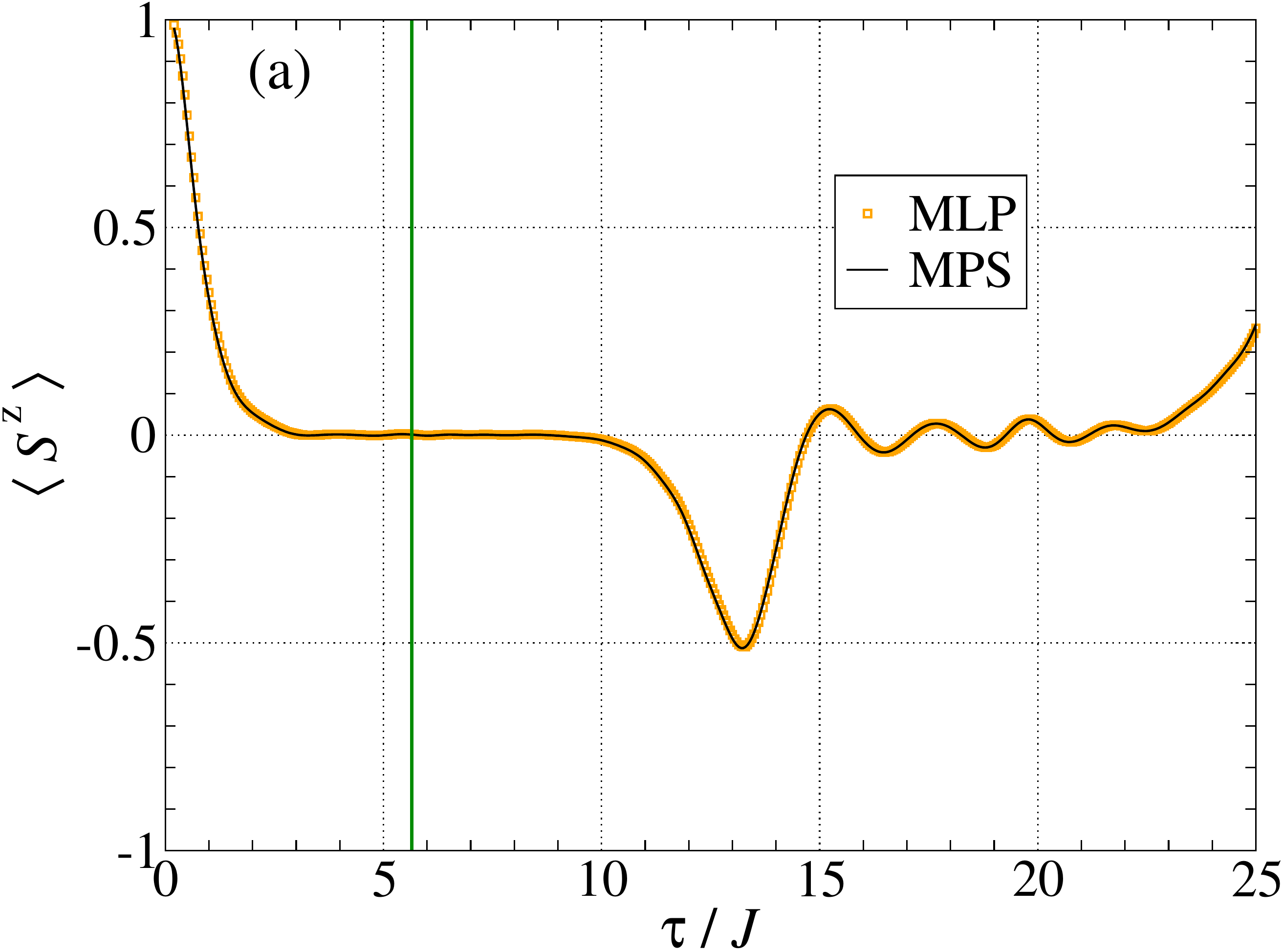}
        }\\
    \centering
    \subfloat[][]{
      \includegraphics[width=0.42\textwidth]{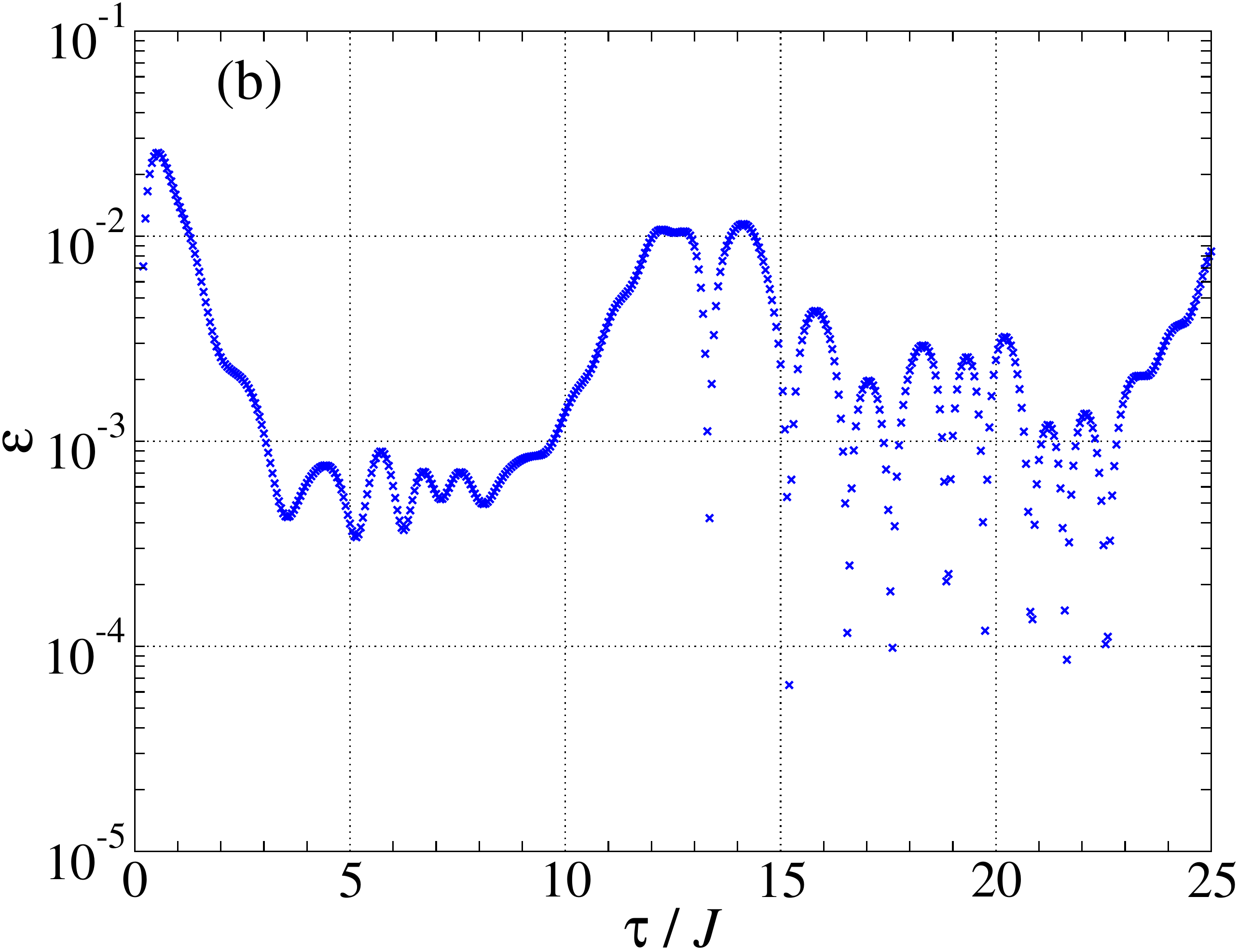}
        }
\caption{(a) Time evolution of the expectation value $\langle S^z
  \rangle$ up to $\tau=25/J$ in time steps of $\delta = 0.05/J$ for
  the one-dimensional Ising spin chain with $N=12$ sites, exchange
  coupling $J$, and transverse field $\Delta = J$ (i.e., at the
  critical point). The MLP used for the regression was constructed
  with 32 linear activated neurons using a training window size of
  $p=4$. The selection of training examples employs the first 110 time
  steps (left of the green line). Results for the MLP regression are
  compared to TEBD results with maximum bond dimension $D=200$. (b)
  The absolute difference $\epsilon = |\langle
  S^z_{\mathrm{MPS}}\rangle - \langle S^z_{\mathrm{MLP}}\rangle|$
  between the TEBD and the MLP regression is shown for each time
  step.}
\label{fig:sz}
\end{figure}

\begin{figure}[tb]
    \centering
    \includegraphics[width=0.35\textwidth]{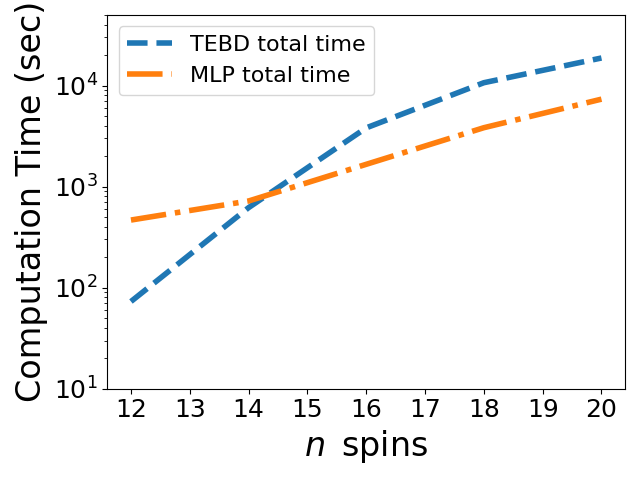}
\caption{A comparison between the total computational times for the
  TEBD and the MLP regression algorithms. Each method was used to
  determine the dynamics of the expectation value of the $S^z$
  operator for the transverse-field Ising model with exchange coupling
  $J$, transverse field, $h=J$, and evolution time $\tau=25/J$ in
  increments of $\delta=0.05/J$. The MLP regression was trained by
  stochastic gradient descent over 32 linearly activated neurons,
  regardless of size. For the TEBD, the maximal bond dimension was set
  to $D=200$ for all sizes.}
\label{fig:computational_time}
\end{figure}

\begin{figure}[tb]
\captionsetup[subfigure]{labelformat=empty}
  \centering
    \subfloat[][]{
      \includegraphics[width=0.42\textwidth]{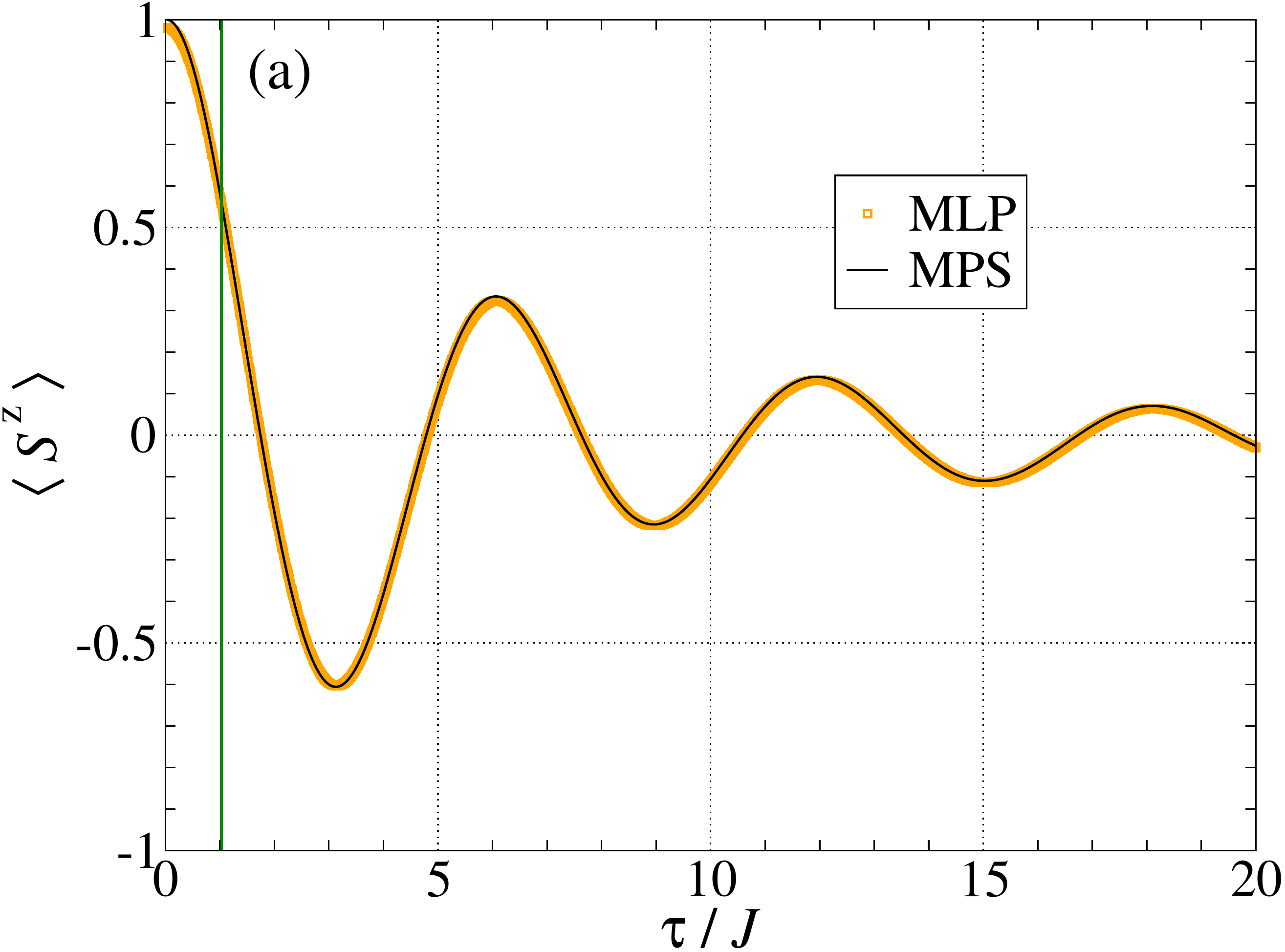}
        }\\
    \centering
    \subfloat[][]{
      \includegraphics[width=0.42\textwidth]{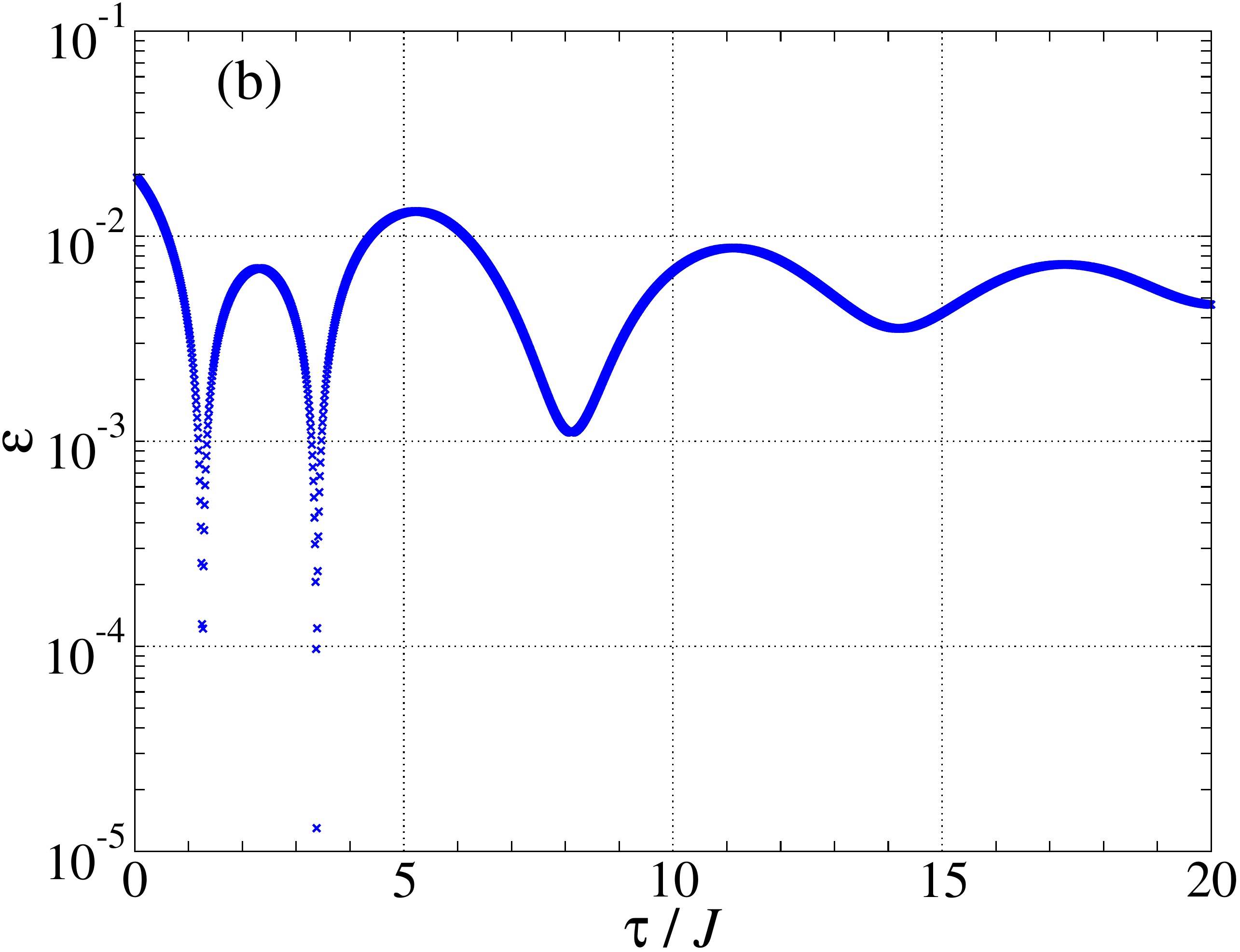}
        }
\caption{(a) Time evolution of the expectation value $\langle S^z
  \rangle$ up to $\tau=20/J$ in times steps of $\delta=0.01/J$ for the
  one-dimensional spin 1/2 XXZ chain with $N=12$ sites, $\Delta =
  J/2$, and $h= J/2$. The MLP used for the regression was constructed
  with 64 linear activated neurons using a training window size of
  $p=4$. The selection of training examples come only from the first
  100 time steps (green line). Results for the MLP regression are
  compared with TEBD results without bond truncation (i.e.,
  exact). (b) The absolute difference $\epsilon = |\langle
  S^z_{\mathrm{MPS}}\rangle - \langle S^z_{\mathrm{MLP}}\rangle|$
  between TEBD calculations and the MLP regression are shown for each
  time step.}
\label{fig:sx}
\end{figure}

We test our MLP regression by evaluating operator expectation values
over two model systems and comparing them to second-order
Trotter-Suzuki time evolved MPS calculations. Firstly, we determine
values at the critical point for the one-dimensional Ising model in a
transverse field with an evolution time of $\tau = 25/J$, where $J$ is
the exchange coupling constant. To compare the computational cost
between the TEBD and the MLP regression, we measure the computational
time as a function of system size. Secondly, we apply the MLP
regression to the one-dimensional XXZ model as a demonstration of its
adaptability to various models. All machine learning simulations were
implemented using the Tensorflow Keras library.\cite{chollet2015}

\subsection{Ising Model}

For $N$ spins in a one-dimensional chain, the nearest-neighbor Ising
model in the presence of a transverse field is given by the
Hamiltonian
\begin{equation}
\mathcal{H} = -J\sum_{i=1}^{N-1} S^z_{i}\, S^z_{i+1} - h
\sum_{i=}^N S^x_{i},
\end{equation}
where $S^z$ is the longitudinal spin operator, $J$ is the exchange
coupling constant, and $h$ characterizes the strength of the
transverse field. Due to the non-commutability of terms in the
Hamiltonian, this model is known to have a quantum phase transition at
$J = h$ in one spatial dimension. This phase transition takes the
system from the ordered ferromagnetic state to a paramagnetic
state.\cite{Strecka2015}

As an illustration of our method, we investigate the dynamics of the
expectation of the local spin operator $\langle S^z \rangle$ near this
phase transition for a short spin chain with $N=12$ spins. We first
use calculations obtained from an MPS with OBC initialized in the
ferromagnetic state to generate expectation values for time steps of
$\delta = 0.05/J$. We split these time-ordered expectation values into
subsets for training and testing the MLP. For our model, we select
training windows of $p=4$, giving us access to 995 input-output
pairs. Of this, we use 110 pairs for training.  Our MLP architecture
is optimized with the following parameters: one layer of 32 linear
activated neurons, followed by a single layer with one linear
activated neuron for output. Training is carried out by a stochastic
gradient descent over the cost function given in
Eq.~(\ref{eq:cost}). Figure~\ref{fig:sz} shows the results with $h =
J$ and maximal bond dimension $D=200$. Within the training region, the
MLP is trained until it has significant overlap with the MPS
calculations. This overlap is seen to continue far past the region of
training. Comparison with the MPS results, as shown in
Fig.~\ref{fig:sz}(b), reveal that the MLP deviates from the MPS
calculations with an average absolute deviation $\epsilon = |\langle
S^z_{\mathrm{MPS}}\rangle - \langle S^z_{\mathrm{MLP}}\rangle|$ equal
to $3\times 10^{-3}$. We note that the training time and parameters
were selected in such a way as to mitigate overfitting for the given
number of training examples, which explains why the deviation is
relatively high in the training range. Using a standard desktop
computer, the time to train the MLP was 415.24 seconds, while the time
to predict the rest of the dynamics was 35.87 seconds. Comparatively,
at this system size, exact diagonalization calculations took
approximately 60 seconds, and TEBD calculations (with fixed bond
dimension $D=200$) took approximately 0.146 seconds per time step,
resulting in a total computation time of 73 seconds.

To glean information about the scaling of the computational cost of
our approach (demonstrating its advantage for larger systems), we
measure the time required to generate short-time TEBD expectation
values as input data and add this to the computational time required
for training and predicting in the long-time regime for varying system
sizes $N$. This computation time is compared to the total time taken
by the TEBD to calculate expectation values over the full time
interval $\tau = 25/J$.  Figure~\ref{fig:computational_time} displays
this comparison. It is clear that the scaling of the computational
time is more favorable for our method. Still, a more interesting
result appears if the time required for generating the input data is
excluded. As shown in Table~\ref{tab:time}, within this regime of
system sizes (having trained until an average deviation of $\epsilon =
10^{-3}$ is achieved), the scaling of the overall computational time
for the MLP regression is due primarily to the time necessary to
generate input-output training pairs. The computational time necessary
for the training and prediction steps in the algorithm appears
polynomial (nearly linear), being primarily due to the necessary
increase in the number of training required to maintain the given
deviation $\epsilon$.

\begin{widetext}

\begin{table}[ht]
\label{tab:time}
\centering
\begin{tabular}{|c|c|c|c|}
\hline
System size & Number of Training Sets Needed & Training Set Generation & Training + Prediction\\
 ($N$) & ($N_{\rm train}$) &(seconds) & (seconds)\\ 
\hline
\hline
12 & 110 & 16.06 & 451.11 \\
14 & 120 & 148.8 &   572.47\\
16 & 140 & 1,069.32 &  594.62 \\
18 & 150 & 3,205.5 & 626.69 \\
20 & 175 & 6,562.5 & 783.55 \\
\hline
\end{tabular}
\caption{Dependence of computational times on systems sizes for the
  transverse-field Ising model. The second column shows the number of
  training examples generated to maintain an average deviation
  $\epsilon = 10^{-3}$. The third column shows the computational times
  to generate the training set of TEBD expectation values. The fourth
  column shows the computational times required for the training and
  predicting stages of the MLP regression.}
\end{table}

\end{widetext}

\subsection{XXZ Model}

We test another ubiquitous spin system with our MLP regression,
namely, the XXZ model. The Hamiltonian governing the evolution of this
open boundary system is given by
\begin{eqnarray}
\mathcal{H} & = & - J\sum_{i=1}^{N-1} \left( S^x_{i} S^x_{i+1} +
S^y_{i} S^y_{i+1} + \Delta\, S^z_{i} S^z_{i+1} \right) \nonumber \\ &
& -\ h \sum_{i=1}^N S^x_i,
\end{eqnarray}
where $J$ and $\Delta$ control the strength of the exchange coupling
and the uniaxial anisotropy, respectively, and $h$ is the strength of
transverse field. The transverse and longitudinal exchange couplings
are $J_{\perp} = J$ and $J_z = J\, \Delta$, respectively. Similar to
the Ising model above, the XXZ model exhibits transitions between the
paramagnetic and the ferromagnetic phases ($J>0$), with critical
values at $h_c = \pm (J_{\perp} - J_z) = J
(1-\Delta)$.\cite{Franchini2017} We investigate this model for $N =
12$ spins and $\Delta = h = J/2$ (within the paramagnetic
phase). Initially, the system is set at the fully-polarized
ferromagnetic state.

We again select a training window of $p=4$, producing 1995
input-output pairs. From these, we train over 100 pairs. The MLP is
composed of a single layer of 64 linearly activated neurons, followed
by an output layer with a single linearly activated neuron. This model
is again trained used stochastic gradient descent. Comparing with the
results taken from MPS calculations over time intervals $\delta =
0.01$, we see in Fig.~\ref{fig:sx} that the MLP regression agrees with
the MPS and continues to do so deep into the testing regime. As shown
in Fig.~\ref{fig:sx}(b), the MLP on average differs consistently from
the TEBD calculations by an average absolute differnce equal to
$6\times 10^{-3}$. The time to sufficiently train to the desired
accuracy was 200.99 seconds, while the time to predict the rest of the
dynamics was 150.14 seconds. Comparatively, exact diagonalization
calculations took approximately 60 seconds, and TEBD calculations
took approximately 300 seconds per time step at the maximal bond
dimension, for a total computation time of approximately 166.67
hours. As for the case of the Ising model, for such a small system,
exact diagonalization is the most cost effective method for computing
$\langle S^z\rangle$, but the cost of this method increases
exponentially with system size as $O(2^{3N})$. As previously shown in
Table~\ref{tab:time}, the scaling is more favorable for MLP.

\section{Discussion}
\label{sec:discussion}

By investigating the evolution of the expectation value of operators,
we have demonstrated that MLP regression accurately extends
calculations in a highly reduced parameter space using very few
training examples. To understand the significance of this, a
comparison between the computational resources used in TEBD and MLP
calculations is presented. For TEBD calculations, long-time dynamics
are obtained by determining the state of the system
$\ket{\Psi_{\mathrm{MPS}}}$ at every time step.\cite{Paeckel2019} For
$N$ sites with maximal bond dimension $D$, this results in using
$O(poly(N)poly(D))$ steps, more specifically, $O(2w^3ND^3)$ steps for
the sequential application of one- and two-body operators, where $w$
is the matrix dimension of the applied local operator.\cite{Orus2014}
Computations with this complexity quickly become cumbersome for long
times, particularly when the correlation length of the system diverges
and the bond dimension $D$ scales exponentially. However, the MLP
regression circumvents this computational cost by utilizing a small
fixed "memory" of previously generated expectation values as the basis
for extending calculations out to long times. As can be seen in
Table~\ref{tab:time}, the computational cost of the combined training
and prediction steps of the regression is approximately polynomial. To
understand how this short "memory" reduces the complexity, consider a
training set having $N_{\mathrm{train}}$ examples constructed over
training windows (i.e., "memories") of size $p$ input into a neural
network have $m$ neurons. For a given value of $p$ elements, the
training phase of the MLP regression has a computational cost which is
determined entirely by the neural network model parameters, $O(m^2p^2
N_{\mathrm{train}})$. After the training, prediction for later times
only has a computational cost of $O(1)$. By generating the first few
expectation values with the MPS, the MLP regression is shown to be
able to predict long-time operator expectations values with only the
addition of a relatively small number of compute cycles. We conclude
that within the regime of sizes considered in this study, the
computational cost of the MLP regression scales remarkably slowly
(nearly constant).

Though this computational advantage is significant, it is worth noting
that the MLP regression can only extend the operator expectation
values generated by the MPS. It is not a generative model and
therefore cannot calculate operator dynamics without the presence of
some initial expectation values. Further work must be done to explore
machine learning architectures which can directly generate operator
dynamics while maintaining low computational costs. Nonetheless, the
results of this work indicate that machine learning techniques
continue to provide unforeseen advantages in modeling QMB systems.

\vspace{-.5cm}

\section*{Acknowledgements}
\label{sec:acknowledgements}

The authors acknowledge partial financial support from NSF grant
No. CCF-1844434.






\end{document}